\documentclass{article}
\usepackage[T1]{fontenc} 
\usepackage[utf8]{inputenc} 
\usepackage{ismir,amsmath,cite,url}
\usepackage{graphicx}
\usepackage{xcolor}
\usepackage[bookmarks=false]{hyperref}
\hypersetup{
    colorlinks=true,
    linkcolor=violet,
    filecolor=magenta,      
    urlcolor=blue,
    citecolor=teal
}
\usepackage{amssymb}
\usepackage{bm}
\usepackage{booktabs}
\usepackage{microtype}
\usepackage{adjustbox}
\usepackage{makecell}
\usepackage{amsmath}
\usepackage{comment}
\usepackage{caption}
\usepackage{lineno}
\usepackage{lipsum} 
\usepackage{booktabs}
\usepackage{enumitem}
\usepackage{multirow}
\usepackage[normalem]{ulem}
\useunder{\uline}{\ul}{}
\usepackage{graphicx}
\usepackage{subcaption} 
\usepackage{hyperref}

\title{Audio Prompt Adapter:\\
Unleashing Music Editing Abilities for Text-to-Music 
with Lightweight Finetuning
}

\multauthor
{Fang-Duo Tsai$^1$ \quad\quad Shih-Lun Wu$^2$ \quad\quad Haven Kim$^3$} {\bfseries{Bo-Yu Chen$^1$ \quad\quad Hao-Chung Cheng$^1$ \quad\quad Yi-Hsuan Yang$^1$} \\
  $^1$\,National Taiwan University \,\,
  $^2$\,Carnegie Mellon University \,\, 
  $^3$\,University of California San Diego \\
  \small{\texttt{r12942150@ntu.edu.tw, shihlunw@andrew.cmu.edu, khaven@ucsd.edu} }
  \\ 
  \small{\texttt{bernie40916@gmail.com, haochung@ntu.edu.tw, yhyangtw@ntu.edu.tw}
  }
}

\def\authorname{F.Tsai, S.Wu, H.Kim, B.Chen, H.Cheng, Y.Yang}

\newcommand{\eric}[1]{\textcolor{black}{#1}}
\newcommand{\fundwo}[1]{\textcolor{black}{#1}}

\begin{document}

\maketitle
%



\begin{abstract}
Text-to-music models allow users to generate nearly realistic musical audio with textual commands.
However, \emph{editing} music audios remains challenging due to the conflicting desiderata of performing fine-grained alterations on the audio while maintaining a simple user interface.
To address this challenge, we propose \textit{Audio Prompt Adapter} (or AP-Adapter), 
a lightweight addition to pretrained text-to-music models.
We utilize AudioMAE to extract features from the input audio,
and construct attention-based adapters to
feed these features into the internal layers of AudioLDM2, a diffusion-based text-to-music model.
With 
22M trainable parameters, AP-Adapter empowers users to harness both global (e.g., genre and timbre) and local (e.g., melody) aspects of music, using the original audio and a short text as inputs.
Through objective and subjective studies, we evaluate AP-Adapter on three tasks: timbre transfer, genre transfer, and accompaniment generation.
Additionally, we demonstrate its effectiveness on out-of-domain audios containing unseen instruments during training.

\end{abstract}

\section{Introduction}\label{sec:introduction}

Advancements in \emph{text-to-music generation} have made it possible for users to create music audio signals from simple textual descriptions~\cite{forsgren2022riffusion,agostinelli2023musiclm,liu2023audioldm,huang2023noise2music}.
To improve the control over the generated music beyond textual input, several newer models have been proposed, using additional conditioning signals indicating the intended global or time-varying musical attributes such as melody, chord progression, rhythm, or loudness for generation \cite{copet2024simple,lin2023content,melechovsky2023mustango,wu2023music,musicongen24ismir} (see Section \ref{sec:related} for a brief review). Such controllability is important for musicians, practitioners, as well as common users in the human-AI co-creation process \cite{HuangKNDC20,louie22iui}.

However, one area that remains challenging, which we refer to as \emph{text-to-music editing} below, is the precise editing of a piece of music, provided by a user as an \emph{audio input} $\bm{x}$ alongside the \emph{text input} $\bm{y}$ for the textual prompts.
The goal here for the model is to create an ``edited'' version of the input music, denoted as $\Tilde{\bm{x}}$, according to the text input.
This is a crucial capability for users who wish to refine either an original or machine-generated music without compromising its musicality and audio quality, while keeping the simplicity of text-based human-computer interaction.
Namely, the desired properties of the output $\Tilde{\bm{x}}$ are:
\begin{itemize}[leftmargin=*,topsep=3pt,itemsep=1pt]
    \item \textbf{Transferability}: $\Tilde{\bm{x}}$ should reflect what $\bm{y}$ specifies, e.g., timbre, genre, instrumentation, or mood.
    \item \textbf{Fidelity}: $\Tilde{\bm{x}}$ should retain all other musical content in $\bm{x}$ that $\bm{y}$ does not concern, e.g., melody and rhythm.
\end{itemize}

\begin{figure*}
 \centerline{
\includegraphics[width=0.9\textwidth]{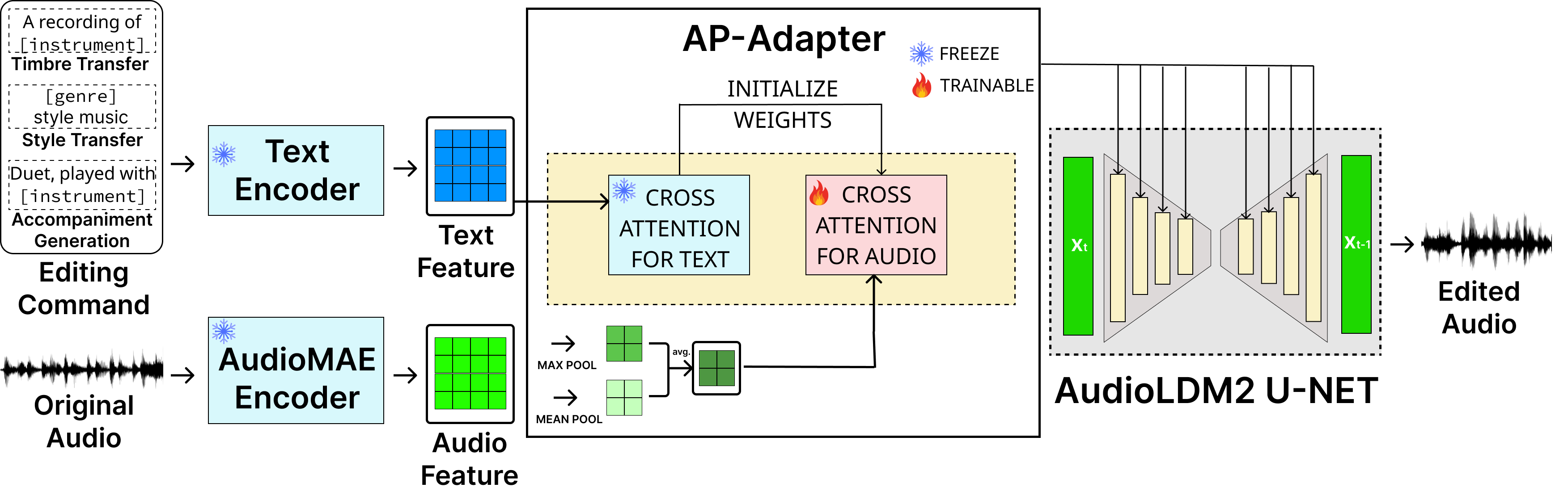}}
 \caption{Our AP-Adapter is an add-on to AudioLDM2~\cite{liu2023audioldm2}.
 Users provide an original audio to AudioMAE~\cite{huang2022masked} to extract audio features, and an editing command to the text encoder.
 The decoupled audio and text cross-attention layers of AP-Adapter contribute to the \textbf{fidelity} with the input audio and \textbf{transferability} of the editing command in the edited audio.
 }
 \label{fig:architecture}
\end{figure*}

While a text-to-music generation model takes in general only the text input $\bm{y}$ and generates music freely, a text-to-music editing model takes both audio and text inputs $\bm{x}$ and $\bm{y}$.
The primary challenge arises from the conflicting goals of maintaining high fidelity to the input audio $\bm{x}$ while incorporating specific changes dictated by textual commands $\bm{y}$.
As we review in Section~\ref{sec:related}, existing methods \cite{han2023instructme,hussain2023m,zhang2024musicmagus} either lack the granularity needed for detailed audio manipulation or need complex prompt engineering that detracts from user accessibility or requires iterative refinements.

A secondary challenge arises from the large number of trainable parameters needed for models to achieve high musical quality and diversity (e.g., MusicGen-medium \cite{copet2024simple} has 1.5B parameters). 
Without much computational resource, it is more feasible to treat existing models as ``foundation models'' and finetune them to fulfill specific needs, instead of training a model from scratch \cite{plitsis2024investigating}. 



In view of these challenges, we propose in this paper the \emph{Audio Prompt Adapter} (or, AP-Adapter for short), a novel approach inspired by the Image Prompt Adapter (IP-Adapter)~\cite{ye2023ip} from the neighboring field of text-to-image editing. This lightweight (22M parameters), attention-based module integrates seamlessly with existing text-to-music generation models, specifically leveraging the pre-trained AudioLDM2 model~\cite{liu2023audioldm2} enhanced by the AudioMAE encoder~\cite{huang2022masked} to extract  audio features. Our method uniquely combines text and audio inputs through decoupled cross-attention layers, allowing precise control in the  generation process. After training the AP-adapter with a single NVIDIA RTX 3090, our method can zero-shot edit a given audio prompt according to the text prompt.

Our AP-Adapter offers great improvements over some baseline models by enabling detailed and context-sensitive audio manipulations, achieving a balance between fidelity and the transferability effects dictated by user inputs.
Our experiments across timbre transfer, genre transfer, and accompaniment generation tasks demonstrate the effectiveness of our approach in handling diverse and complex editing requirements.
In short, our key contributions are:
\begin{itemize}[leftmargin=*,topsep=3pt,itemsep=1pt]
    \item Proposing a framework that equips an audio input modality for a pre-trained text-to-music generation model.
    \item Performing zero-shot music editing with a lightweight adapter, which permits flexible balance of the effects of the text and audio inputs.
    \item Demonstrating three tasks: timbre transfer, genre transfer, accompaniment generation, and discussing the impact of tunable hyperparameters. 
\end{itemize}
We provide audio examples in our demo website.\footnote{Demo: \url{https://rebrand.ly/AP-adapter}
} We also share source code and model checkpoints on GitHub.\footnote{Code: \url{https://github.com/fundwotsai2001/AP-adapter}}

%


\section{Related Work}
\label{sec:related}


Generating desired music from text prompts alone is complex and often requires intricate prompt engineering.  Mustango~\cite{melechovsky2023mustango} enhanced prompts with information-rich captions specifying chords, beats, tempo, and key. MusicGen~\cite{copet2024simple} conditioned music generation on melodies by extracting chroma features~\cite{ellis2007chroma} and inputting them with the text prompt into a Transformer model. 
Coco-Mulla~\cite{lin2023content} and \eric{MusiConGen} \cite{musicongen24ismir} extended MusicGen by adding time-varying chord- and rhythm-related controls.
Music ControlNet~\cite{wu2023music} incorporated time-varying conditions like melody, rhythm, and dynamics for diffusion-based text-to-music models.  These methods utilize low-level features to guide generation but do not take reference audio as input, limiting their potential for editing existing audio tracks.

Recently, several music editing methods were proposed. InstructME~\cite{han2023instructme} uses a VAE and a chord-conditioned diffusion model for music editing but requires a large dataset of audio files with multiple instrumental tracks and triplet data of text instructions, source music, and target music for supervised training. M$^2$UGen~\cite{hussain2023m} leverages large language models to understand and generate music across different modalities, supporting music editing via natural language, but it requires a three-step training process and complex preprocessing. MusicMagus~\cite{zhang2024musicmagus} implements latent space manipulation during inference for music editing but requires an additional music captioning model and the InstructGPT LLM to address discrepancies between the text prompt distribution of AudioLDM2 and the music captioning model.

Compared to these methods, our AP-Adapter is more straightforward to train and can achieve multiple music editing tasks in a zero-shot manner.

\section{Background}\label{sec:bg}
\subsection{Diffusion Model}\label{subsubsec:diffusion}
Denoising diffusion probabilistic models (DDPMs)~\cite{ho2020denoising}, also known as diffusion models, are a class of generative models that approximates some distribution $p(\bm{x})$ via denoising through a sequence of $T-1$ latent variables:
\begin{equation}
    p_{\theta}(\bm{x}) = \int \Big[\prod_{t=1}^{T}p_{\theta}(\bm{x}_{t-1} \, | \, \bm{x}_t) \Big] p(\bm{x}_T) d\bm{x}_{1:T} \, ,
\end{equation}
where $\theta$ is the set of learnable parameters, $\bm{x}_0 := \bm{x}$, and $p(\bm{x}_T) := \mathcal{N}(0, \mathbf{I})$ (i.e., an uninformative Gaussian prior).
To train the model, we run forward diffusion: sample some data point $\bm{x} \sim p(\bm{x})$ and some $t \in [1, T]$, and add noise $\bm{\epsilon} \sim \mathcal{N}(0, \mathbf{I})$ to $\bm{x}$ to produce a noised data point $\bm{x}_t := \sqrt{\bar{\beta}_t} \bm{x} + \sqrt{1 - \bar{\beta}_t} \bm{\epsilon}$, where $\bar{\beta}_t$ is the pre-defined noise level for step $t$.
The model is asked to perform backward diffusion, namely, to recover the added noise via the objective $ \min_{\theta} \mathbb{E}_{\bm{x},\bm{\epsilon},t}\left[ \|\bm{\epsilon} - \bm{\epsilon}_{\theta}(\bm{x}_t, t)\|_2^2 \right]$, where $\bm{\epsilon}_{\theta}(\cdot)$ is the model's prediction, that is equivalent to maximizing the evidence lower bound (ELBO) of $p_{\theta}(\bm{x})$.
During inference, we start from an $\bm{x}_T \sim \mathcal{N}(0, \mathbf{I})$ and iteratively remove the predicted noise $\bm{\epsilon}_{\theta}(\bm{x}_t, t)$ to generate data.
Song \emph{et al.}~\cite{song2020score} offered a crucial interpretation that each denoising step can be seen as ascending along $\nabla_{\bm{x}} \log p_{\theta}(\bm{x})$, also known as the \textit{score} of $p_{\theta}(\bm{x})$.
Any input condition $\bm{y}$ can be incorporated into a diffusion model by injecting embeddings of $\bm{y}$ via, for example, cross-attention~\cite{rombach2022high},
thereby modeling $p_{\theta}(\bm{x} \,|\,\bm{y})$ (and $\nabla_{\bm{x}} \log p_{\theta}(\bm{x} \,|\, \bm{y})$).
To reduce memory footprint and accelerate training/inference, latent diffusion models (LDMs)~\cite{rombach2022high} proposed to first compress data points $\bm{x}$ into latent vectors using a variational autoencoder (VAE)~\cite{kingma2013auto}, and then learn a diffusion model for the latent vectors.

\subsection{AudioLDM2}\label{subsubsec:audioldm2}
We choose AudioLDM2~\cite{liu2023audioldm2}, a latent diffusion-based~\cite{rombach2022high} text-to-audio model, as our pretrained backbone.
To enable text control over generated audio, AudioLDM2 uses AudioMAE~\cite{huang2022masked} to extract acoustic features, named the \textit{language of audio} (LOA), from the target audio.
LOA serves as the bridge between acoustic and text-centric semantic information---the text prompt is encoded by both the FLAN-T5~\cite{chung2024scaling} language model and CLAP~\cite{laionclap2023} text encoder (which has a joint audio-text embedding space), and then passed to a trainable GPT-2~\cite{radford2019language} to approximate the LOA via a regression loss that aligns the semantic representations with LOA.
The aligned text information is then fed into the U-Net~\cite{ronneberger2015u} for diffusion process to influence the generation.
We pick AudioLDM2 to be the backbone since the use of LOA likely promotes the affinity to accepting audio conditions, which is crucial to our fidelity goal.


\subsection{Classifier-free Guidance}\label{subsec:cfg}
Classifier-free guidance (CFG)~\cite{ho2022classifier} is a simple yet effective inference-time method to enhance the input text condition's influence, which is directly linked to our transferability goal.
As mentioned in Sec.~\ref{subsubsec:diffusion}, diffusion models can predict both the unconditioned score $\nabla_{\bm{x}} \log p(\bm{x})$ and the conditioned score $\nabla_{\bm{x}} \log p(\bm{x} \mid \bm{y})$.
In addition, by Bayes' rule, we know that $p(\bm{x} \mid \bm{y}) \propto p(\bm{x})p(\bm{y} \mid \bm{x})$.
As the goal is the amplify $\bm{y}$'s influence, we define:
\begin{equation}
    p_{\lambda}(\bm{x} \mid \bm{y}) :\propto p(\bm{x})p(\bm{y} \mid \bm{x})^{\lambda} \, , 
\end{equation}
where $\lambda$ is a knob, named \textit{CFG scale}, that controls the strength of $\bm{y}$.
Taking $(\nabla_{\bm{x}}\log)$ on both sides 
gives us:
\begin{equation}\label{eqn:cfg-knob}
    \nabla_{\bm{x}} \log p_{\lambda}(\bm{x} \mid \bm{y}) = \lambda \nabla_{\bm{x}} \log p(\bm{y} \mid \bm{x}) + \nabla_{\bm{x}} \log p(\bm{x}) \, .
\end{equation}
Meanwhile, we can rearrange the Bayes' rule terms to get:
\begin{equation}\label{eqn:bayes-reuse}
    \nabla_{\bm{x}} \log p(\bm{y} \mid \bm{x}) = \nabla_{\bm{x}}\log p(\bm{x} \mid \bm{y}) - \nabla_{\bm{x}} \log p(\bm{x}) \, .
\end{equation}
Note that a diffusion model can predict both RHS terms.
Plugging Eqn.~\eqref{eqn:bayes-reuse} into Eqn.~\eqref{eqn:cfg-knob}, CFG performs
\begin{align}\label{cfg}
  &\nabla_{\bm{x}} \log p_{\lambda}(\bm{x} \mid \bm{y}) = \nabla_{\bm{x}} \log p(\bm{x})\nonumber \\
& \quad\quad + \lambda(\nabla_{\bm{x}} \log p(\bm{x} \mid \bm{y}) -  \nabla_{\bm{x}} \log p(\bm{x}))
\end{align}
at every inference iteration, 
where $\nabla_{\bm{x}} \log p(\bm{x})$ is obtained by inputting an empty string as $\bm{y}$.

\section{Proposed Audio Prompt Adapter} 

To effectively condition AudioLDM2 on the input audio and achieve our transferability and fidelity goals,
our AP-Adapter adds two components to AudioLDM2: an audio encoder to extract acoustic features, and decoupled cross-attention adapters to incorporate the acoustic features while maintaining text conditioning capability.

\subsection{Audio Encoder and Feature Pooling}\label{subsec:audio-encoder}
We adopt AudioMAE as the audio encoder, which is used by AudioLDM2 to produce the language of audio (LOA; see Section~\ref{subsubsec:audioldm2}) during its training.
In our pilot study,
we find that using the LOA directly as the condition causes nearly verbatim reconstruction, i.e., information in the input audio is mostly retained. This is undesirable as it greatly limits transferability.
To address this issue, we apply a combination of max and mean pooling on the LOA, and leave the pooling rate, which we denote by $\omega$, tunable by the user to trade off between fidelity and transferability.


\subsection{Decoupled Cross-attention Adapters}\label{subsec:decouple}
According to the analyses in~\cite{gal2023encoder, kumari2023multi} performed on text-to-image diffusion models finetuned for image editing~\cite{huggingface2024sd-dreambooth}, the cross-attention layers, which allow interaction between text prompt and the diffusion process, undergo the most drastic changes during fine-tuning.
Hence, we implement our AP-Adapter also as a set of cross-attention layers.

Recall that the audio and text prompts are transformed to internal features before interacting with the U-Net for diffusion. We define these features as:
\begin{align}
    \bm{c}_{\bm{x}} &:= \text{Pool}(\text{AudioMAE}(\bm{x})) \label{eqn:c_x} \\
    \bm{c}_{\bm{y}} &:= \text{GPT2}([\text{FlanT5}(\bm{y}); \text{CLAP}(\bm{y})]) \, , \label{eqn:c_y} 
\end{align}
where $\bm{c}_{\bm{x}}$ and $\bm{c}_{\bm{y}}$ are the audio and text features respectively.
The original AudioLDM2 incorporates the text feature into each U-Net layer via cross-attention:
\begin{equation}\label{eqn:text-cross-attn}
\bm{z}_{\text{text}} := \text{Attention}(\bm{z}\bm{W}^{(q)}, \bm{c}_{\bm{y}}\bm{W}^{(k)}, \bm{c}_{\bm{y}}\bm{W}^{(v)}) \, ,
\end{equation}
where $\bm{z}$ is the U-Net's internal feature, and $\bm{W}^{(q)}$, $\bm{W}^{(k)}$, $\bm{W}^{(v)}$ are learnable projections that respectively produce the cross-attention query, key, and values from $\bm{z}$ or $\bm{c}_{\bm{y}}$.
We keep this cross-attention for text intact (i.e., frozen), anticipating it to satisfy transferability out of the box.

To incorporate the audio features for fidelity, we place a decoupled audio cross-attention layer as the adapter alongside each text cross-attention in a similar light to~\cite{ye2023ip}:
\begin{equation}\label{eqn:audio-cross-attn}
\bm{z}_{\text{audio}} := \text{Attention}(\bm{z}\bm{W}^{(q)}, \bm{c}_{\bm{x}}\bm{W'}^{(k)}, \bm{c}_{\bm{x}}\bm{W'}^{(v)}) \, ,
\end{equation}
where $\bm{W'^{(k)}}$ and $\bm{W'^{(v)}}$ are the newly introduced adapter weights.
Since during AudioLDM2 training, the text feature $\bm{c}_{\bm{y}}$ is trained to mimic the LOA from AudioMAE, we initialize $\bm{W'^{(k)}}$ and $\bm{W'^{(v)}}$ respectively from $\bm{W^{(k)}}$ and $\bm{W^{(v)}}$ for all the cross-attention layers in the Unet, and find that this significantly shortens our fine-tuning process compared to random initialization.

Finally, we obtain the final output of the decoupled text and audio cross-attentions via a weighted sum:
\begin{equation}\label{eqn:attn-sum}
\bm{z}_{\text{fusion}} := \bm{z}_{\text{text}} + \alpha \bm{z}_{\text{audio}} \, ,
\end{equation}
where $\alpha \in \mathbb{R}$, named \textit{AP scale}, is a hyperparameter that controls the strength of the audio prompt (fixed to $\alpha=1$ during training), and $\bm{z}_{\text{fusion}}$ becomes the input of the subsequent U-Net layer.
We expect $\bm{z}_{\text{fusion}}$ to capture the information mixture from audio and text prompts, inducing the model to generate plausible music that adheres to both.

\subsection{Training} 
We freeze all the parameters in the pretrained AudioLDM2 and AudioMAE, except for the decoupled audio cross-attention adapters with 22M parameters.
The loss function follows that of standard (latent) diffusion models:
\begin{equation}
   \mathcal{L} = \mathbb{E}_{(\bm{x}, \bm{y}),\bm{\epsilon},t} \left\| \bm{\epsilon} - \bm{\epsilon}_{\theta}\left(\bm{x}_t, \bm{c}_{\bm{x}}, \bm{c}_{\bm{y}}, t\right) \right\|^2_2 \, ,
   \label{eqn:loss function}
\end{equation}
where $(\bm{x}, \bm{y})$ are naturally existing paired audio and text, $\bm{\epsilon} \sim \mathcal{N}(0,\mathbf{I})$, $t$ is the diffusion step, $\bm{x}_t$ is the noised audio latent features, $\bm{c}_{\bm{x}}, \bm{c}_{\bm{y}}$ are the extracted features from text and audio prompts (cf. Eqn.~\eqref{eqn:c_x} and~\eqref{eqn:c_y}), and $\bm{\epsilon}_{\theta}(\cdot)$ is the model's predicted noise.
Minimizing $\mathcal{L}$ is equivalent to maximizing the lower bound of $p(\bm{x} \mid \bm{c}_{\bm{x}}, \bm{c}_{\bm{y}})$.
During training, we select the audio feature's pooling rate $\omega$ from the set $\{1, 2, 4, 8\}$ uniformly at random, making the adapters recognize audio features with different resolutions, thereby allowing users to balance fidelity and transferability at inference. 
Additionally, we randomly dropout audio and text conditions, i.e., setting $\bm{c}_{\bm{x}}$ to a zero matrix, and $\bm{y}$ to an empty string, to facilitate classifier-free guidance.

\subsection{Inference}
At inference, users are free to input any text prompt $\bm{y}$ as the editing command to achieve their desired edits, i.e, $\bm{x}\rightarrow\Tilde{\bm{x}}$.
In addition, following~\cite{StableDiffusion2024NegativePrompt, sanchez2023stay}, we modify the unconditioned terms in Eqn.~\eqref{cfg} using a negative text prompt $\bm{y}^{-}$.
Letting $\bm{c}_{\bm{xy}} := \{\bm{c}_{\bm{x}}, \bm{c}_{\bm{y}}\}$, our inference step
is:
\begin{align} \label{negative}
 &\nabla_{\Tilde{\bm{x}}} \log p_{\lambda}(\Tilde{\bm{x}} \,|\, \bm{c}_{\bm{xy}}, \bm{c}_{\bm{y^{-}}}) = \nabla_{\Tilde{\bm{x}}} \log p(\Tilde{\bm{x}} \,|\,\bm{c}_{\bm{y^{-}}}) \nonumber \\
 & \quad\quad + \lambda \left(\nabla_{\Tilde{\bm{x}}} \log p(\Tilde{\bm{x}} \,|\,\bm{c}_{\bm{xy}}) - \nabla_{\Tilde{\bm{x}}} \log p(\Tilde{\bm{x}} \,|\,\bm{c}_{\bm{y^{-}}})\right)
\end{align}
We find that specifying $\bm{y^{-}}$ is an effective way to avoid unwanted properties in $\Tilde{\bm{x}}$, e.g., the original timbre for the timbre transfer task, or low-quality music in general.
%

\section{Experiment Setup}\label{sec:experiments}

\subsection{Dataset Preparation}\label{subsec:body1}
For the training data of our AP-Adapter, due to our limited computation resource, we use 200K 10-second-long audios with text tags randomly 
sampled from AudioSet~\cite{gemmeke2017audio} (about 500 hours, or $\sim$10\% of the whole dataset).

\eric{For the audio input $\bm{x}$ used in evaluation}, we compile two datasets: \textbf{in-domain} and \textbf{out-of-domain}, according to whether the AudioSet ontology includes the instrument.
\begin{itemize}[leftmargin=*,itemsep=0pt,topsep=2pt]
    \item \textbf{In-domain}: We choose 8 common instruments: \texttt{piano}, \texttt{violin}, \texttt{cello}, \texttt{flute}, \texttt{marimba}, \texttt{organ}, \texttt{harp} and \texttt{acoustic guitar}.
    For each instrument, we manually download 5 high-quality monophonic audios from YouTube (i.e., 40 samples in total) and crop them each to 10 seconds.
    \item \textbf{Out-of-domain}: We collect a dataset of monophonic melodies played by ethnic instruments,
    including 2 \emph{Chinese} instruments (collected by one of our co-authors) and 5 \emph{Korean} instruments (downloaded from AIHub~\cite{aihubdataset}).
    We use 5 audio samples for each instrument (35 audios in total), cropped to 10 seconds each. \eric{We note that these instruments are \textbf{not seen} during the training time.}
\end{itemize}
\eric{Except for the Korean data which is not licensed outside of Korea, we share information to get the data on GitHub.}

\subsection{Evaluation Tasks}\label{subsec:expr-tasks}

By varying the edit command $\bm{y}$, we evaluate AP-Adapter on three music editing tasks:
\begin{itemize}[leftmargin=*,itemsep=0pt,topsep=2pt]
    \item \textbf{Timbre transfer}: The model is expected to change a melody's timbre to that of the target instrument, and keep all other contents unchanged. 
    For this task, the editing command ($\bm{y}$) is set to ``a recording of a \texttt{[target instrument]} solo''.
    The negative prompt ($\bm{y^{-}}$) is ``a recording of the \texttt{[original instrument]} solo''.
    \eric{For in-domain input, the target is one of the other 7 in-domain instruments. For out-of-domain input, the target is one of the 8 in-domain instruments. We only use in-domain instruments as the target because our evaluation metrics CLAP~\cite{laionclap2023} and FAD~\cite{kilgour2018fr} (see Section \ref{subsec:obj-metrics}) do not recognize the out-of-domain instruments.} 
    \item \textbf{Genre transfer}: We expect the genre (e.g., jazz and country) to change according to the text prompt, but we wish to retain most of the other content such as melody, rhythm and timbre.
    Here, we set $\bm{y} :=$ ``\texttt{[target genre]} style music'', and $\bm{y^{-}} :=$ ``low quality music''. \eric{Here, we target 8 genres: \texttt{jazz}, \texttt{reggae}, \texttt{rock}, \texttt{metal}, \texttt{pop}, \texttt{hip-hop}, \texttt{disco}, \texttt{country}.}
    \item \textbf{Accompaniment generation}: We expect that all content in the input melody remains unchanged, but a new instrument is added to accompany the original audio in a pleasant-sounding and harmonic way.
    We set $\bm{y} :=$ ``Duet, played with \texttt{[accomp instrument]} accompaniment'', and $\bm{y^{-}} :=$ ``low quality music''. \eric{The [accomp instrument] is selected in the same way as the [target instrument] in the timbre transfer task.}
\end{itemize}
We include these representative tasks which musicians may find useful for their daily workflow, but since $\bm{y}$ is free-form text, AP-Adapter has the potential for many other tasks.



\begin{figure*}[htbp!]
  \centering
  \begin{subfigure}[b]{0.33\textwidth}
    \includegraphics[width=\textwidth]{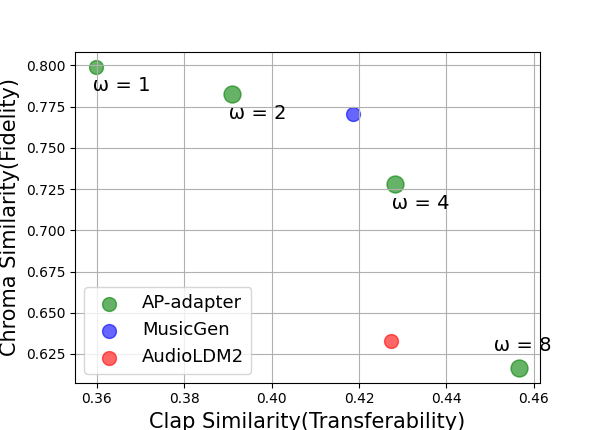}
    \caption{Tuning pooling rate $\omega$}
    \label{fig:pooling}
  \end{subfigure}
  \hfill 
  \begin{subfigure}[b]{0.33\textwidth}
    \includegraphics[width=\textwidth]{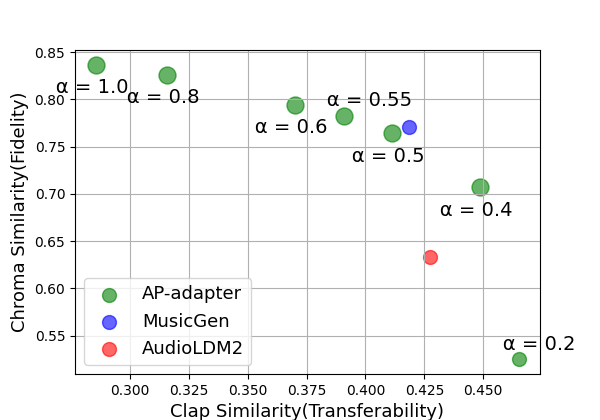}
    \caption{Tuning AP scale $\alpha$}
    \label{fig:Ap}
  \end{subfigure}
  \hfill 
  \begin{subfigure}[b]{0.33\textwidth}
    \includegraphics[width=\textwidth]{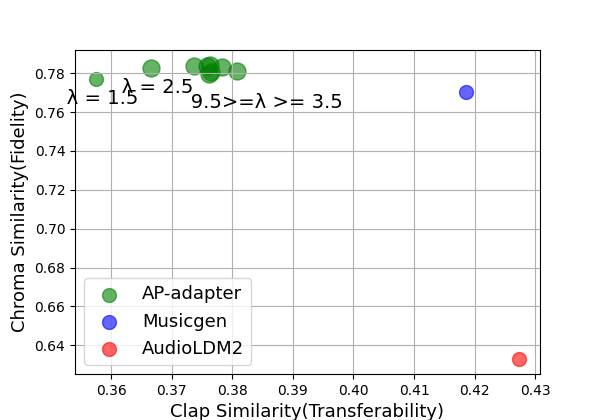}
    \caption{Tuning classifier-free guidance scale $\lambda$}
    \label{fig:cfg}
  \end{subfigure}
  \vspace{-4mm}
 \caption{
    Transferability-fidelity tradeoff effects of different hyperparameters on the timbre transfer task.
    The hyperparameters are set to $\omega$ = 2, $\alpha$ = 0.55, and $\lambda$ = 7.5 when they are not the hyperparameter of interest.
    }
  \label{fig:pareto}
\end{figure*}

\subsection{Training and Inference Specifics}\label{subsec:body2}
We use AudioLDM2-large (1.5B parameters), available on HuggingFace, as our backbone model, and only train our 22M-parameter adapters.
Training is done on a single
one RTX 3090 (24GB) for 35K steps with an effective batch size of 32.
We use AdamW optimizer with fixed learning rate $10^{-4}$ and weight decay $10^{-2}$.
To enable CFG, we randomly dropout text and audio features with a 5\% probability.

For inference, we choose the critical hyperparameters, i.e., pooling rate $\omega$, AP scale $\alpha$, and CFG scale $\lambda$, by 
exploring the transferability-fidelity tradeoff space as will be reported in Section~\ref{subsec:pareto}.
For timbre transfer and accompaniment generation, we select $\omega$ = 2, $\alpha$ = 0.5, $\lambda$ = 7.5.
For the genre transfer , we select $\omega$ = 1, $\alpha$ = 0.4, $\lambda$ = 7.5.
Following AudioLDM2, we use 50 diffusion steps.

\subsection{Baselines}\label{subsec:body3}
We choose two well-known 
and publicly-available 
audio generation models, AudioLDM2~\cite{liu2023audioldm2} and MusicGen~\cite{copet2024simple}, as our baselines. Both of them can generate nearly realistic music.
We describe below how we use them for editing: 
\begin{itemize}[leftmargin=*,itemsep=0pt,topsep=2pt]
    \item \textbf{AudioLDM2}: Following \eric{SDEdit}~\cite{meng2021sdedit}, we perform the forward process (i.e., adding noise to the audio input $\bm{x}$) partially for $0.75T$ steps, where $T$ is the original number diffusion steps, and then denoise it back with the editing command $\bm{y}$ to obtain $\Tilde{\bm{x}}$.
    \item \textbf{MusicGen}:
    MusicGen is a Transformer-based text-to-audio model that generates discrete audio tokens.
    We use MusicGen-Melody (1.5B), which achieves melody conditioning using 
    chromagram~\cite{ellis2007chroma} as a proxy.
    We input $\bm{y}$ as the text prompt, and the chromagram of $\bm{x}$ as the audio condition, for MusicGen to generate $\Tilde{\bm{x}}$.
\end{itemize}
We do not include the recent text-to-music editing methods InstructME~\cite{han2023instructme} or MusicMagus~\cite{zhang2024musicmagus}, as they have not released the code and models, and also exclude M$^2$UGen~\cite{hussain2023m}  as it is heavily focused on music understanding and visually-conditioned music generation.

\subsection{Objective Metrics}\label{subsec:obj-metrics}
We employ the following metrics:
\begin{itemize}[leftmargin=*,itemsep=0pt,topsep=2pt]
    \item \textbf{CLAP}~\cite{laionclap2023} is used to evaluate \textbf{transferability}, as it is trained with contrastive losses to align the representations for audio and text.
    We compute the cosine similarity between CLAP audio embedding for the edited audio $\Tilde{\bm{x}}$ and CLAP text embedding for the  command $\bm{y}$.\footnote{For accompaniment generation task, text input to CLAP is modified to include both instruments, e.g., ``Piano duet, played with violin.''}
    Higher scores show high semantic relevance between  $\Tilde{\bm{x}}$ and $\bm{y}$.
    \item \textbf{Chroma similarity} computes the similarity of the original and edited audios $\Tilde{\bm{x}}$ and $\bm{x}$ harmonically and rhythmically, thereby evaluates \textbf{fidelity}.  We adopt librosa's~\cite{mcfee2015librosa} CQT chroma method to extract the 12-dimensional chromagrams~\cite{ellis2007chroma} 
    to compute framewise cosine similarity.
    \item \textbf{Fr\'echet audio distance (FAD)}~\cite{kilgour2018fr} uses a pretrained audio classifier to extract audio features, collects features from all audios, and estimates the feature covariance matrix.
    Then, the Fr\'echet distance is computed between the two covariance matrices (one from generated audios, one from real audios).
    We adopt FAD to evaluate the \textbf{overall quality/realisticness} of the generations.
    Following the official implementation, we use VGGish architecture~\cite{hershey2017cnn} as the feature extractor.
    We use the in-domain evaluation dataset as real audios. 
    
\end{itemize}

\subsection{Subjective Study}\label{subsec:subj-study}
We design a listening test that contains 2 sets of music for each of the three tasks.
The sets are independent from one another, and each contains a 10-second original audio prompt $\bm{x}$, an editing text command $\bm{y}$, and three edited audios $\Tilde{\bm{x}}$ generated by our model and the two baselines (with order randomized and kept secret to participants).
Participants rate each edited audio on a 5-point Likert scale, according to the following \eric{3} aspects:
\begin{itemize}[leftmargin=*,noitemsep,nolistsep]
    \item \textbf{Transferability}: Do you feel that the generated audio matches what the text prompt asks for?
    \item \textbf{Fidelity}: Do you feel that the generated audio faithfully keeps the original musical content that should not be changed by the text prompt?
    \item \textbf{Overall preference}: Overall, how much do you like the generated audio?
\end{itemize}
We recruit 30 participants from our social circle and randomly assign them one of the 6 test suites (3 for in-domain, 3 for out-of-domain).
The study takes about \eric{10} minutes.

\begin{table}[t]
\centering
\begin{minipage}{0.46\textwidth}
\large 
\begin{adjustbox}{max width=\textwidth}
\begin{tabular}{lccc}
\toprule
\textbf{Model}  & \makecell{\textbf{CLAP}\,$\uparrow$ \\ (transferability)} & \makecell{\textbf{Chroma}\,$\uparrow$ \\ (fidelity)} & \makecell{\textbf{FAD}\,$\downarrow$ \\ (overall)} \\
\midrule
MusicGen & \textbf{0.339} & 0.771 & 8.443 \\
AudioLDM2  & 0.284 & 0.643 & \textbf{5.389} \\
AP-Adapter & 0.314 & \textbf{0.777} & 5.986 \\
\bottomrule
\end{tabular}
\end{adjustbox}
\caption{Objective evaluation on \emph{in-domain} audio inputs of \eric{MusicGen-Melody~\cite{copet2024simple}, AudioLDM2-SDEdit~\cite{liu2023audioldm2,meng2021sdedit}, and the proposed AP-Adapter.} \fundwo{Results are the average of the three tasks.} Best results are highlighted in bold ($\uparrow$\,/\,$\downarrow$: the higher\,/\,lower the better).}
\label{tab:obj}
\end{minipage}
\end{table}


\begin{table*}[th]
\centering{
\resizebox{0.99\textwidth}{!}{%
\begin{tabular}{llccccccccccc}
\hline
\multirow{2}{*}{} & \multicolumn{1}{r}{\textit{Metric}} & \multicolumn{3}{c}{\textbf{Transferability \eric{MOS}}} & \multicolumn{3}{c}{\textbf{Fidelity \eric{MOS}}} & \multicolumn{3}{c}{\textbf{Overall \eric{MOS}}} \\ \cmidrule(lr){3-5} \cmidrule(lr){6-8} \cmidrule(lr){9-11}
 \textit{Eval. audios} & \multicolumn{1}{r}{\textit{Task}} & Timbre & Genre & Accomp. & Timbre & Genre & Accomp. & Timbre & Genre & Accomp. \\ \hline
\multirow{3}{*}{In-domain} & MusicGen & 3.35 & 3.15 & 3.32 & 2.62 & 2.85 & 2.76 & 3.06 & 3.03 & 2.91 \\
 & AudioLDM2 & 3.21 & 2.74 & 3.12 & 2.21 & 2.21 & 2.26 & 2.47 & 2.56 & 2.47 \\
 & AP-Adapter & \textbf{3.59} & \textbf{3.44} & \textbf{3.41} & \textbf{3.47} & \textbf{3.74} & \textbf{3.41} & \textbf{3.26} & \textbf{3.44} & \textbf{3.12} \\ \hline
\multirow{3}{*}{Out-of-domain} & MusicGen & \textbf{2.92} & \textbf{3.96} & 3.00 & 2.73 & 3.31 & 2.54 & 2.58 & \textbf{3.58} & 2.65 \\
 & AudioLDM2 & 2.62 & 2.12 & 2.96 & 2.42 & 2.69 & 2.23 & 2.58 & 2.31 & 2.81 \\
 & AP-Adapter & \textbf{2.92} & 3.19 & \textbf{3.54} & \textbf{3.81} & \textbf{3.58} & \textbf{3.96} & \textbf{3.08} & 3.12 & \textbf{3.31} \\ \hline
\end{tabular}
}
}
\vspace{-1mm}
\caption{Subjective study results (mean opinion scores \eric{$\in [1,5]$}) with 17 and 13 participants for in-domain and out-of-domain input audios, respectively, \eric{for the three evaluation tasks: timbre transfer, genre transfer, and accompaniment generation}. 
} 
\label{survey}
\vspace{-2mm}
\end{table*}

\section{Results and Discussion}\label{sec:typeset_text}
\subsection{Hyperparameter Choices}\label{subsec:pareto}
We discover in our early experiments that several hyperparameters, which are tunable during inference,
can drastically affect the edited outputs.
Therefore, we conduct a systematic study on the effects of audio pooling rate $\omega$ (Sec.~\ref{subsec:audio-encoder}), AP scale in decoupled cross-attention $\alpha$ (Sec.~\ref{subsec:decouple}), and classifier-free guidance scale $\lambda$ (Sec.~\ref{subsec:cfg}).
Specifically, we observe how their various values induce different behaviors on the transferability-fidelity plane spanned by CLAP and chroma similarity metrics.
\begin{itemize}[leftmargin=*,itemsep=0pt,topsep=1pt]
    \item The \textbf{pooling rate} $\omega$ controls the amount of information from the audio prompt. Figure~\ref{fig:pooling} shows clearly that when the pooling rate is low, the fidelity is higher, but at the cost of transferability. For example, the audio generated with $\omega=1$ preserves  abundant acoustic information, thus the edited audio sounds like the input audio, but it might not reflect the editing command. 
    The opposite can be said for  $\omega=8$.
    Overall, $\omega=2$ or $4$ strikes a good balance.
    \item The \textbf{AP scale} $\alpha$ adjusts the relative importance between the text and audio decoupled cross-attentions.
    As opposed to pooling rate, it enhances fidelity at the expense of transferability at higher values, as shown in Figure~\ref{fig:Ap}, and $\alpha \in [0.4, 0.6]$ leads to a more balanced performance.
    
    \item  The \textbf{CFG guidance scale}  $\lambda$
    dictates the strength of text condition as detailed in Eqn.~(\ref{cfg}).
    As shown in Figure~\ref{fig:cfg}, somewhat unexpectedly, $\lambda$ does not impact the tradeoff too much when $\lambda \geq 3.5$.
    Hence, we use $\lambda = 7.5$ across all tasks following AudioLDM2.
\end{itemize}

\subsection{Objective Evaluations}\label{subsec:body4}
We show the metrics computed on in-domain audios 
in Table \ref{tab:obj}, \eric{taking the average across the three editing tasks}. 
(We do not report the result for out-of-domain audio inputs as we expect CLAP and FAD to be less reliable there.)
In general, AP-Adapter exhibits the most well-rounded performance without significant weaknesses---MusicGen scores high on transferability, but has a much worse FAD score, indicating issues on quality or distributional deviation. \fundwo{We infer that, since MusicGen only considers melody as input rather than the entire audio, it has fewer limitations in the generating process and thus achieves a higher transferability score.}
\fundwo{On the other hand, AudioLDM2 consistently achieves the best FAD score but lacks fidelity and transferability.}

We also evaluate the ablated version of  AP-Adapter without using the negative prompt ($\bm{y^{-}}$).
For the timber transfer task, not using the negative prompt induces worse transferability, degrading the CLAP score from 0.405 to 0.378, but does not negatively impact chroma similarity and FAD.

\subsection{Subjective Evaluations}\label{subsec:body5}

Table~\ref{survey} shows the results from our listening test.
Our AP-adapter outperforms the two other baseline models in 16 out of 18 comparisons.
On top of preserving fine-grained
details in the input audio, 
AP-adapter also tightly follows the editing commands and generate relatively high-quality music, leading in transferability and overall preference except for only the genre transfer task on out-of-domain audios.
\eric{MusicGen performs better in transferability for genre transfer, but its fidelity is weaker as it only considers the melody of the input audio.}
With the additional audio-modality condition, AP-adapter has the advantage of ``listening'' to all the details of the input audio, receiving significantly higher scores on fidelity on both in- and out-of-domain cases. 

\eric{The advantage of AP-adapter in fidelity is much stronger in Table~\ref{survey} rahter than in Table~\ref{tab:obj}.}
We conjecture that chroma similarity paints only a partial picture for fidelity as it is focused primarily on harmonic properties, leaving out other musical elements such as dynamics and percussive patterns.

\section{Conclusions}
\label{sec:conclusion}
We presented AP-Adapter, a lightweight add-on to AudioLDM2 that empowers it for music editing.
AP-Adapter leverages AudioMAE to extract fine-grained features from the audio prompt, and feeds such features into AudioLDM2 via decoupled cross-attention adapters for effective conditioning.
With only 500 hours of training data and 22M trainable parameters, AP-Adapter delivers compelling performance across useful editing tasks, namely, timbre transfer, genre transfer, and accompaniment generation.
Additionally, it enables users to manipulate the transferability-fidelity tradeoff, and edit out-of-domain audios, which promotes creative endeavors with ethnic instrument audios that are usually scarce in publicly available datasets.

Promising directions for follow-up works include: \textit{(i)}~exploring more diverse editing tasks under our framework with various editing commands, 
\textit{(ii)}~extending AP-Adapter to other generative backbones, e.g., autoregressive models, and
\textit{(iii)}~adding support for localized edits that can be stitched seamlessly with unchanged audio segments.



\section{Acknowledgment}

The work is also partially supported by a grant from the National Science and
Technology Council of Taiwan (NSTC 112-2222-E-002-005-MY2) and (NSTC 113-2628-E-002 -029), and Ministry of Education (NTU-112V1904-5).

\bibliography{ISMIRtemplate}

\begin{thebibliography}{10}
\providecommand{\url}[1]{#1}
\csname url@samestyle\endcsname
\providecommand{\newblock}{\relax}
\providecommand{\bibinfo}[2]{#2}
\providecommand{\BIBentrySTDinterwordspacing}{\spaceskip=0pt\relax}
\providecommand{\BIBentryALTinterwordstretchfactor}{4}
\providecommand{\BIBentryALTinterwordspacing}{\spaceskip=\fontdimen2\font plus
\BIBentryALTinterwordstretchfactor\fontdimen3\font minus \fontdimen4\font\relax}
\providecommand{\BIBforeignlanguage}[2]{{%
\expandafter\ifx\csname l@#1\endcsname\relax
\typeout{** WARNING: IEEEtran.bst: No hyphenation pattern has been}%
\typeout{** loaded for the language `#1'. Using the pattern for}%
\typeout{** the default language instead.}%
\else
\language=\csname l@#1\endcsname
\fi
#2}}
\providecommand{\BIBdecl}{\relax}
\BIBdecl

\bibitem{forsgren2022riffusion}
\BIBentryALTinterwordspacing
S.~Forsgren and H.~Martiros, ``{Riffusion: Stable diffusion for real-time music generation},'' 2022. [Online]. Available: \url{https://riffusion.com}
\BIBentrySTDinterwordspacing

\bibitem{agostinelli2023musiclm}
A.~Agostinelli, T.~I. Denk, Z.~Borsos, J.~Engel, M.~Verzetti, A.~Caillon, Q.~Huang, A.~Jansen, A.~Roberts, M.~Tagliasacchi \emph{et~al.}, ``{Music{LM}}: Generating music from text,'' \emph{arXiv preprint arXiv:2301.11325}, 2023.

\bibitem{liu2023audioldm}
H.~Liu, Z.~Chen, Y.~Yuan, X.~Mei, X.~Liu, D.~Mandic, W.~Wang, and M.~D. Plumbley, ``{Audio{LDM}}: Text-to-audio generation with latent diffusion models,'' in \emph{Proceedings of International Conference on Machine Learning (ICML)}, 2023.

\bibitem{huang2023noise2music}
Q.~Huang, D.~S. Park, T.~Wang, T.~I. Denk, A.~Ly, N.~Chen, Z.~Zhang, Z.~Zhang, J.~Yu, C.~Frank, J.~Engel, Q.~V. Le, W.~Chan, Z.~Chen, and W.~Han, ``Noise2music: Text-conditioned music generation with diffusion models,'' \emph{arXiv preprint arXiv:2302.03917}, 2023.

\bibitem{copet2024simple}
J.~Copet, F.~Kreuk, I.~Gat, T.~Remez, D.~Kant, G.~Synnaeve, Y.~Adi, and A.~D{\'e}fossez, ``Simple and controllable music generation,'' \emph{Advances in Neural Information Processing Systems (NeurIPS)}, 2024.

\bibitem{lin2023content}
L.~Lin, G.~Xia, J.~Jiang, and Y.~Zhang, ``Content-based controls for music large language modeling,'' \emph{arXiv preprint arXiv:2310.17162}, 2023.

\bibitem{melechovsky2023mustango}
J.~Melechovsky, Z.~Guo, D.~Ghosal, N.~Majumder, D.~Herremans, and S.~Poria, ``Mustango: Toward controllable text-to-music generation,'' in \emph{Proceedings of Conference of the North American Chapter of the Association for Computational Linguistics (NAACL)}, 2024.

\bibitem{wu2023music}
S.-L. Wu, C.~Donahue, S.~Watanabe, and N.~J. Bryan, ``{Music ControlNet}: Multiple time-varying controls for music generation,'' \emph{IEEE/ACM Transactions on Audio, Speech, and Language Processing}, vol.~32, pp. 2692--2703, 2024.

\bibitem{musicongen24ismir}
Y.-H. Lan, W.-Y. Hsiao, H.-C. Cheng, and Y.-H. Yang, ``{MusiConGen}: Rhythm and chord control for {Transformer}-based text-to-music generation,'' in \emph{International Society for Music Information Retrieval Conference (ISMIR)}, 2024.

\bibitem{HuangKNDC20}
C.~A. Huang, H.~V. Koops, E.~Newton{-}Rex, M.~Dinculescu, and C.~J. Cai, ``{Human-AI} co-creation in songwriting,'' in \emph{International Society for Music Information Retrieval Conference (ISMIR)}, 2020.

\bibitem{louie22iui}
R.~Louie, J.~H. Engel, and C.~A. Huang, ``Expressive communication: {A} common framework for evaluating developments in generative models and steering interfaces,'' in \emph{ACM Intelligent User Interfaces Conference (IUI)}, 2022.

\bibitem{liu2023audioldm2}
H.~Liu, Q.~Tian, Y.~Yuan, X.~Liu, X.~Mei, Q.~Kong, Y.~Wang, W.~Wang, Y.~Wang, and M.~D. Plumbley, ``Audio{LDM} 2: Learning holistic audio generation with self-supervised pretraining,'' \emph{IEEE/ACM Transactions on Audio, Speech, and Language Processing}, vol.~32, pp. 2871--2883, 2024.

\bibitem{huang2022masked}
P.-Y. Huang, H.~Xu, J.~Li, A.~Baevski, M.~Auli, W.~Galuba, F.~Metze, and C.~Feichtenhofer, ``Masked autoencoders that listen,'' \emph{Advances in Neural Information Processing Systems (NeurIPS)}, 2022.

\bibitem{han2023instructme}
B.~Han, J.~Dai, X.~Song, W.~Hao, X.~He, D.~Guo, J.~Chen, Y.~Wang, and Y.~Qian, ``{InstructME}: An instruction guided music edit and remix framework with latent diffusion models,'' \emph{arXiv preprint arXiv:2308.14360}, 2023.

\bibitem{hussain2023m}
A.~S. Hussain, S.~Liu, C.~Sun, and Y.~Shan, ``{M$^2$UGen}: Multi-modal music understanding and generation with the power of large language models,'' \emph{arXiv preprint arXiv:2311.11255}, 2023.

\bibitem{zhang2024musicmagus}
Y.~Zhang, Y.~Ikemiya, G.~Xia, N.~Murata, M.~Mart{\'\i}nez, W.-H. Liao, Y.~Mitsufuji, and S.~Dixon, ``{MusicMagus}: Zero-shot text-to-music editing via diffusion models,'' in \emph{Proceedings of International Joint Conferences on Artificial Intelligence (IJCAI)}, 2024.

\bibitem{plitsis2024investigating}
M.~Plitsis, T.~Kouzelis, G.~Paraskevopoulos, V.~Katsouros, and Y.~Panagakis, ``Investigating personalization methods in text to music generation,'' in \emph{IEEE International Conference on Acoustics, Speech and Signal Processing (ICASSP)}, 2024, pp. 1081--1085.

\bibitem{ye2023ip}
H.~Ye, J.~Zhang, S.~Liu, X.~Han, and W.~Yang, ``{IP-Adapter}: Text compatible image prompt adapter for text-to-image diffusion models,'' \emph{arXiv preprint arXiv:2308.06721}, 2023.

\bibitem{ellis2007chroma}
D.~Ellis, ``Chroma feature analysis and synthesis,'' \emph{Resources of laboratory for the recognition and organization of speech and Audio-LabROSA}, 2007.

\bibitem{ho2020denoising}
J.~Ho, A.~Jain, and P.~Abbeel, ``Denoising diffusion probabilistic models,'' \emph{Advances in Neural Information Processing Systems (NeurIPS)}, 2020.

\bibitem{song2020score}
Y.~Song, J.~Sohl-Dickstein, D.~P. Kingma, A.~Kumar, S.~Ermon, and B.~Poole, ``Score-based generative modeling through stochastic differential equations,'' in \emph{International Conference on Learning Representations (ICLR)}, 2021.

\bibitem{rombach2022high}
R.~Rombach, A.~Blattmann, D.~Lorenz, P.~Esser, and B.~Ommer, ``High-resolution image synthesis with latent diffusion models,'' in \emph{Proceedings of the IEEE/CVF Conference on Computer Vision and Pattern Recognition (CVPR)}, 2022.

\bibitem{kingma2013auto}
D.~P. Kingma and M.~Welling, ``Auto-encoding variational bayes,'' \emph{arXiv preprint arXiv:1312.6114}, 2013.

\bibitem{chung2024scaling}
H.~W. Chung, L.~Hou, S.~Longpre, B.~Zoph, Y.~Tay, W.~Fedus, Y.~Li, X.~Wang, M.~Dehghani, S.~Brahma \emph{et~al.}, ``Scaling instruction-finetuned language models,'' \emph{Journal of Machine Learning Research (JMLR)}, 2024.

\bibitem{laionclap2023}
Y.~Wu, K.~Chen, T.~Zhang, Y.~Hui, T.~Berg-Kirkpatrick, and S.~Dubnov, ``Large-scale contrastive language-audio pretraining with feature fusion and keyword-to-caption augmentation,'' in \emph{IEEE International Conference on Acoustics, Speech and Signal Processing (ICASSP)}, 2023.

\bibitem{radford2019language}
A.~Radford, J.~Wu, R.~Child, D.~Luan, D.~Amodei, and I.~Sutskever, ``Language models are unsupervised multitask learners,'' \emph{Open AI Blog}, 2019.

\bibitem{ronneberger2015u}
O.~Ronneberger, P.~Fischer, and T.~Brox, ``U-net: Convolutional networks for biomedical image segmentation,'' in \emph{International Conference on Medical Image Computing and Computer-Assisted Intervention}, 2015.

\bibitem{ho2022classifier}
J.~Ho and T.~Salimans, ``Classifier-free diffusion guidance,'' \emph{arXiv preprint arXiv:2207.12598}, 2022.

\bibitem{gal2023encoder}
R.~Gal, M.~Arar, Y.~Atzmon, A.~H. Bermano, G.~Chechik, and D.~Cohen-Or, ``Encoder-based domain tuning for fast personalization of text-to-image models,'' \emph{ACM Transactions on Graphics (TOG)}, 2023.

\bibitem{kumari2023multi}
N.~Kumari, B.~Zhang, R.~Zhang, E.~Shechtman, and J.-Y. Zhu, ``Multi-concept customization of text-to-image diffusion,'' in \emph{Proceedings of the IEEE/CVF Conference on Computer Vision and Pattern Recognition (CVPR)}, 2023, pp. 1931--1941.

\bibitem{huggingface2024sd-dreambooth}
H.~concept library, ``{SD-Dreambooth-Library},'' \url{https://huggingface.co/sd-dreambooth-library}, 2024, [Online; accessed 10-April-2024].

\bibitem{StableDiffusion2024NegativePrompt}
\BIBentryALTinterwordspacing
{Stable Diffusion Art}, ``How does negative prompt work?'' 2024, [Online; accessed 10-April-2024]. [Online]. Available: \url{https://stable-diffusion-art.com/how-negative-prompt-work/}
\BIBentrySTDinterwordspacing

\bibitem{sanchez2023stay}
G.~Sanchez, H.~Fan, A.~Spangher, E.~Levi, P.~S. Ammanamanchi, and S.~Biderman, ``Stay on topic with classifier-free guidance,'' \emph{arXiv preprint arXiv:2306.17806}, 2023.

\bibitem{gemmeke2017audio}
J.~F. Gemmeke, D.~P.~W. Ellis, D.~Freedman, A.~Jansen, W.~Lawrence, R.~C. Moore, M.~Plakal, and M.~Ritter, ``{Audio Set}: An ontology and human-labeled dataset for audio events,'' in \emph{IEEE international conference on acoustics, speech and signal processing (ICASSP)}, 2017.

\bibitem{aihubdataset}
``{AI Hub Dataset},'' \url{https://www.aihub.or.kr/aihubdata/data/view.do?currMenu=115&topMenu=100&dataSetSn=71470}.

\bibitem{kilgour2018fr}
K.~Kilgour, M.~Zuluaga, D.~Roblek, and M.~Sharifi, ``{Fr\'echet Audio Distance}: A metric for evaluating music enhancement algorithms,'' \emph{arXiv preprint arXiv:1812.08466}, 2018.

\bibitem{meng2021sdedit}
C.~Meng, Y.~He, Y.~Song, J.~Song, J.~Wu, J.-Y. Zhu, and S.~Ermon, ``{SDEdit}: Guided image synthesis and editing with stochastic differential equations,'' in \emph{International Conference on Learning Representations (ICLR)}, 2022.

\bibitem{mcfee2015librosa}
B.~McFee, C.~Raffel, D.~Liang, D.~P. Ellis, M.~McVicar, E.~Battenberg, and O.~Nieto, ``librosa: Audio and music signal analysis in python.'' in \emph{SciPy}, 2015.

\bibitem{hershey2017cnn}
S.~Hershey, S.~Chaudhuri, D.~P. Ellis, J.~F. Gemmeke, A.~Jansen, R.~C. Moore, M.~Plakal, D.~Platt, R.~A. Saurous, B.~Seybold \emph{et~al.}, ``{CNN} architectures for large-scale audio classification,'' in \emph{IEEE International Conference on Acoustics, Speech and Signal Processing (ICASSP)}, 2017.

\end{thebibliography}
\end{document}